\documentclass[aps,prl,reprint]{revtex4-1}

\usepackage{amsmath}    
\usepackage{graphicx}   
\usepackage{color}      
\usepackage{lmodern} 
\usepackage{gensymb}
\usepackage{pdfpages}
\usepackage{pgffor}

\makeatletter
\AtBeginDocument{\let\LS@rot\@undefined}
\makeatother

\begin{document}
\title{Low-energy phonons in Bi$_2$Sr$_2$CaCu$_2$O$_{8+\delta}$ and their possible interaction with electrons measured by inelastic neutron scattering}

\author{A. M. Merritt$^1$, J.-P. Castellan$^2$, T. Keller $^{3,4}$, S. R. Park$^5$, J. A. Fernandez-Baca$^{6, 7}$, G. D. Gu$^8$, D. Reznik$^{1, 9}*$}

\affiliation{1. Department of Physics, University of Colorado - Boulder, Boulder, Colorado 80309, USA\\2. Karlsruhe Institute of Technology, 76021 Karlsruhe, Germany\\3. Max-Planck-Institut f\"ur Festk\"orperphysik, 70569 Stuttgart, Germany\\4. Max Planck Society Outstation at the FRM II, 85748 Garching, Germany\\5. Department of Physics, Incheon National University, Incheon 22012, Republic of Korea\\6. Neutron Scattering Division, Oak Ridge National Laboratory, Oak Ridge, Tennessee 37831, USA\\7. Department of Physics and Astronomy, University of Tennessee, Knoxville, Tennessee 37996, USA\\8. Condensed Matter Physics \& Materials Science Department, Brookhaven National Laboratory, Upton, New York 11973, USA\\9.  Center for Experiments on Quantum Materials, University of Colorado - Boulder, Boulder, Colorado, 80309, USA\\* Corresponding author: Dmitry.Reznik@colorado.edu}

\begin{abstract}
Electron-phonon interaction in copper oxide superconductors is still enigmatic. Strong coupling for certain optic phonons is now well established experimentally, but  theoretical understanding is challenging. Scattering of electrons near the Fermi surface by the longitudinal acoustic (LA) phonons is expected from basic theory because these phonons modulate electron density. We used inelastic neutron scattering on a large single crystal sample of optimally-doped Bi$_2$Sr$_2$CaCu$_2$O$_{8+\delta}$ to show that low-energy LA phonons could couple to electronic density fluctuations only at small phonon wavevectors, which naturally limits any interaction to forward scattering. Such scattering should not be pairbreaking in the case of the d-wave gap. We also found that previously the reported low energy phonon spectral weight half-way to the zone boundary is consistent with conventional lattice dynamics and does not reflect an incipient charge density wave.

\end{abstract}

\maketitle

\section{Introduction}

In metals phonons can scatter electrons between different parts of the Fermi surface. This scattering can contribute to Cooper pair formation if the superconducting gap amplitudes at the initial and final electron momenta have the same sign. If the signs are opposite, the phonons break Cooper pairs. In d-wave superconductors such as the cuprates, both types of scattering are possible, so the contributions of phonons to pair forming and pair breaking tend to partially cancel. Phonon scattering with small momentum transfer (forward scattering) can be pair forming because it scatters electrons predominantly between nearby regions of the Fermi surface where the superconducting gaps have the same sign \cite{reznik_q_1995,bulut_dx2y2_1996,sandvik_effect_2004}.

Electron phonon interaction is proportional to electron-phonon matrix elements. According to many theoretical models, they are negligibly small  in copper oxide superconductors, however, many experiments show pronounced anomalous softening and broadening of some Cu-O bond-stretching and bond-buckling optic phonons \cite{franck_oxygen_1991, thomsen_raman_1992, shen_role_2002, gadermaier_electron-phonon_2010, reznik_giant_2010, reznik_phonon_2012-1}. These anomalies appear mostly close to mid-way between the zone center and the zone boundary along the Cu-O bond direction (a*). In YBa$_2$Cu$_3$O$_{6+\delta}$ they renormalize strongly below the superconducting transition temperature T$_c$, which is a clear sign of strong electron-phonon interaction \cite{chung_-plane_2003-1,reznik_temperature_2008}. There are indications that they are associated with incipient or actual charge density wave (CDW) charge order with wavevector q$_{\text{co}}$=0.25-0.3 reciprocal lattice units (r.l.u.) induced by low-energy electronic charge fluctuations. These charge fluctuations have been observed by resonant inelastic x-ray scattering (RIXS) in many compounds, especially on the underdoped side of the phase diagram \cite{Ghiringhelli_LongRangeIncommensurateCharge-2012, daSilvaNeto_UbiquitousInterplayCharge-2014, Chaix_Dispersivechargedensity-2017}.

Additional evidence for strong electron-phonon coupling comes from angle-resolved photoemission spectroscopy (ARPES) measurements on cuprates that report pronounced kinks in electronic dispersions near the energies of the anomalous optic phonons (30-80 meV). The origin of these kinks is still not fully settled but some argue that it originates from electron-phonon interaction \cite{cuk_coupling_2004, reznik_e-ph_coupling_ARPES_2008}. In the simplest case of an electronic band coupling to a flat bosonic mode such as an Einstein phonon, electron-phonon interaction is expected to induce such a kink in the electronic dispersion close to the mode energy \cite{ashcroft_solid_1976}. In the superconducting state, the gap at the final electron momentum adds to the kink energy, which is called gap referencing \cite{rameau_10meV_kink_optical_ph_2009,  plumb_10meV_kink_2010}. For example, in a copper oxide superconductor  Bi$_2$Sr$_2$CaCu$_2$O$_{8+\delta}$ (Bi-2212 or BSCCO) with the superconducting gap maximum $\Delta_{max} \sim 30$ meV, an optic phonon branch with a wavevector independent electron-phonon matrix element will induce a kink in the nodal electron dispersion at $\omega_{ph}+\Delta_{max}>30$ meV. The relationship between these kinks and the above-mentioned phonon anomalies has not been established; in fact some of us have argued that there is none, since the CDW is a collective mode not directly seen in ARPES \cite{Park_Evidencechargecollective-2014}.

This work explores low-energy phonons in optimally-doped BSCCO showing that previously identified signatures of incipient CDW are well explained by conventional lattice dynamics.

We also considered the possible phononic origin of the electronic dispersion kinks near 10 meV that have been reported more recently \cite{rameau_10meV_kink_optical_ph_2009,  plumb_10meV_kink_2010, anzai_10meV_kink_2010, vishik_10meV_kink_2010}. These kinks show gap referencing to a local gap rather than the maximum gap, i.e. E$_{kink}$(k)=$\omega_{ph}+\Delta_{k}$, where k is the electron crystal momentum. Referencing to a local gap implies forward scattering so that the wavevector of the final state of the electron-phonon scattering process is close to the initial state. A low-energy phonon branch is a particularly compelling candidate because other bosonic modes in BSCCO, such as spin fluctuations, appear at higher energies. However, an experimental doping-dependent study of the kink in Bi-2201 and theoretical work on Bi-2212 show a significant dependence on doping \cite{kondo_10meV_kink_2201_2013, johnston_acoustic_ph_2012}. Previous work has suggested that a low-energy optical mode could be responsible \cite{rameau_10meV_kink_optical_ph_2009, He_Persistentlowenergyphonon-2018}, but Raman scattering and optical experiments do not see a significant doping dependence of phonon energies, which is inconsistent with this mechanism. Other work has suggested forward scattering from an acoustic phonon, with doping-dependent screening of electron-phonon coupling \cite{johnston_acoustic_ph_2012}.

We used inelastic neutron scattering to map out the dispersions of low energy phonons while searching for possible mechanisms of forward scattering in a large high-quality single-crystal sample of Bi$_2$Sr$_2$CaCu$_2$O$_{8+\delta}$. We found that a longitudinal acoustic (LA) phonon branch is the most natural candidate to explain the 10meV kink. Since these phonons modulate the electronic density, the electron-phonon coupling mechanism is well understood \cite{mahan_gerald_many-particle_2000}. Furthermore, the q-dependence of the phonon eigenvectors confines strong electron-phonon coupling to small phonon wavevectors as shown in Fig. 1d,e.

\begin{figure}[htb!]
\includegraphics[width=\linewidth]{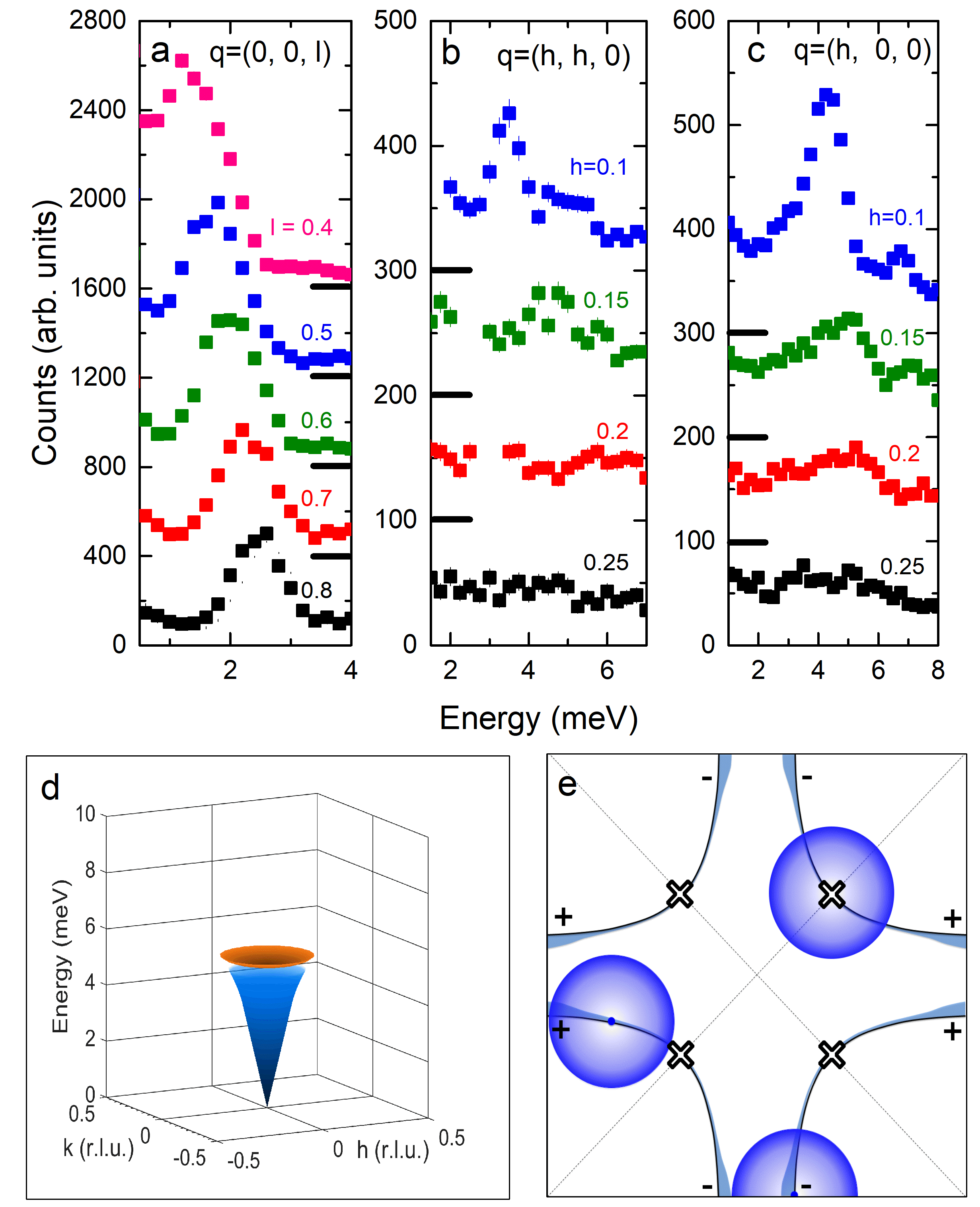}
\caption{LA phonons relevant for the 10 meV ARPES kink. (a) Acoustic phonon dispersing in the l-direction from (0, 0,  12). The peak intensity is similar across the BZ. T=90K, E$_f$=8 meV. Points are offset by 400 counts; the thick black lines denote the baseline for each \textbf{q}. (b, c) LA phonons dispersing along the (110) and (100) directions. Total wavevectors \textbf{Q} are \textbf{Q}=(1,1,0)+\textbf{q} (b) and \textbf{Q}=(2,0,0)+\textbf{q} (c). The intensities decrease rapidly as they move away from the ZC. T=10K, E$_f$=8meV.  Points are offset by 100 counts; the thick black lines denote the baseline for each \textbf{q}. (d) The blue cone/orange disk indicates acoustic/lowest optic phonon dispersion at small wavevectors before mixing, consistent with our measurements. (e) Schematic of the bonding Fermi surface, superconducting gap and electron-phonon scattering in BSCCO. The black curves denote the bonding Fermi surface. The straight lines pass through the four nodes each marked with an X. Kinematic constraints for forward scattering of electrons at the centers of the semitransparent blue circles by LA phonons are met only inside the circles as discussed in the text. The variation in transparency of the circles represents the intensity of the scattering that is proportional to $\lvert \textbf{q} \rvert$.}
\label{fig:Phonon_Schematic_1}
\end{figure}

\section{Methods}

Measurements were performed on the 1T triple-axis spectrometer at the Orph\'{e}e reactor (Laboratoire L\'{e}on Brillouin, France) using a PG002 monochromator/analyzer in the standard open collimation configuration and in a similar condition on the HB-3 spectrometer at the High Flux Isotope Reactor (HFIR) at Oak Ridge National Laboratory. The sample was a large single crystal grown with the floating-zone technique as detailed in Ref. \cite{wen_large_2008}. The sample was mounted in a closed-cycle refrigerator. We looked at every experimentally accessible Brillouin zone (BZ) around each L, \textbf{Q}=(00L), as well as (200) and (202), using fixed final energies E$_f$=8, 13.2, and 14.7 meV. For the majority of the measurements the sample was mounted in the H0L scattering plane (in our notation H is along a*, K is along b* and L is along c*). Results are reported in reciprocal lattice units (r.l.u.) with a=b=3.82 \AA \ and c=30.9 \AA. We measured phonon dispersions along the H and L-directions in reciprocal space as well as along the HH0/H-H0 and some off-symmetry directions. HH0 direction is defined to be parallel to the superlattice modulation wavevector as shown in Fig. S1 (the Fig. S$\#$ scheme will be used to refer to figures in the supplementary information, \cite{supplementary_info}). The energy region around 10 meV in the data measured with $E_f$=8 meV is not used because it corresponds to a $3k_f=2k_i$ condition. 

Neutron scattering intensity of phonons is determined, in part, by their eigenvectors \cite{shirane_neutron_2002}. Acoustic phonon intensities are maximized near the strong Bragg peaks and suppressed around the weak ones, whereas the optic phonons often follow the opposite trend. For acoustic phonons we performed measurements in the Brillouin zones adjacent to the strong Bragg peaks at \textbf{Q}=(2,0,0), \textbf{Q}=(1,1,0) and \textbf{Q}=(0,0,12). Optic phonons were measured in zones adjacent to weak Bragg peaks at \textbf{Q}=(2,0,2) and \textbf{Q}=(0,0,14). See Fig. S2 for a schematic of cuts in the HK0 plane.

\begin{figure*}[htb!]
\includegraphics[width=\textwidth]{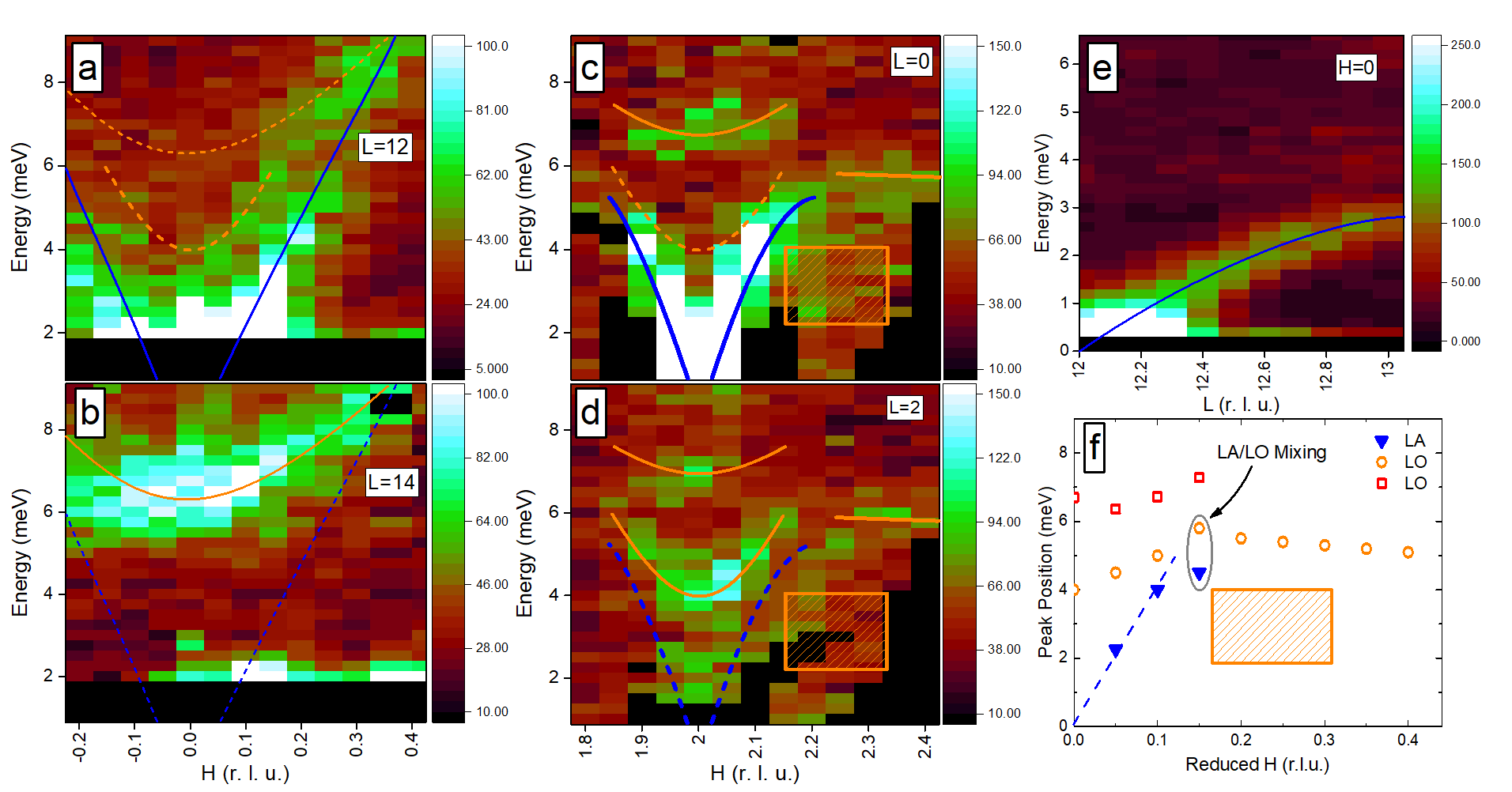}
\caption{Main experimental results at low temperature for Q=(H, 0, L). (a-e) Momentum-energy cuts. Solid lines denote phonon dispersions visible in a zone: blue for acoustic and orange for optic. Dashed lines indicate that the phonon is not visible in that particular BZ, but the location is known from other BZs. The orange rectangle is the region exhibiting superlattice TA phonon peaks. Final neutron energy for (a, b) was E$_f$=14.7 meV at T=10K. (c, d) E$_f$=8 meV T=2K. (e) E$_f$=8 meV, T=90K. (f) Peak positions for phonons dispersing along H. Dashed line indicates a linear dispersion expected of the LA branch (blue triangles) in the absence of interaction with the LO branch (orange circles).}
\label{fig:Combined}
\end{figure*}

The experiment on phonon lifetime was conducted at the spin echo NRSE-TAS spectrometer TRISP  \cite{Keller_2002_TRISP} at the Maier-Leibnitz-Zentrum (MLZ) in Garching, Germany. The BSCCO sample was mounted in a closed-cycle cryostat in exchange gas in the (HK0) scattering plane. The superlattice peaks were observed in the elastic signal to ensure proper orientation. Neutron spin-echo measurements were performed at Q=(2.11,0,0) and Q=(2.12,0,0).

\section{Results}


LA phonons form sound waves in the continuum approximation in which the standard theory can give appreciable electron-phonon coupling \cite{mahan_gerald_many-particle_2000}. Such phonons are strong only near intense Bragg peaks and their intensity evolves as 1/$\omega$(q) at low temperature. This is true for the LA branch along c* as illustrated in Fig. 1a, where its intensity does not change appreciably throughout the zone. On the other hand, the in-plane LA-phonon intensity drops rapidly beyond h=0.1 r.l.u. (Fig. 1b,c and Fig. S3). 

Figures 2,3 provide insight into the origin of this behavior. Figs. 2a,b shows a typical low-energy transverse phonon spectrum near an intense Q=(0,0,12) Bragg peak and near a weak one with Q=(0,0,14). Here the reduced wavevector is perpendicular to the total wavevector, which selects TA phonons dispersing along H (i.e. a*) and atomic displacements along the c-axis (parallel to Q). In Fig. 2a the spectrum is dominated by the TA branch indicated by the solid blue line. The TO phonon dispersion deduced from the other Brillouin zone is indicated by the orange dashed line, but the intensity is small. In Fig. 2b the intensity of the acoustic phonons is negligibly small  (hence the blue line is dashed), but the TO phonon indicated by a solid orange line is clearly visible. This difference between phonon intensities in the two Brillouin zones is entirely due to the different phonon structure factors.

Fig. 2c,d and Fig. 3a-e contrast the behavior in a Brillouin zone next to the (2,0,0) Bragg peak that emphasizes the LA branch and another zone, near (2,0,2), that emphasizes the LO branches. Both branches disperse along a*. The LA branch disperses steeply upwards, whereas the LO branches have a much more gradual dispersion. As the LA branch approaches the LO branch away from the zone center, the LA and LO characters can be distinguished by the different intensities in the two Brillouin zones. For example, at h=0.1 the LA phonon is strong at Q=(2.1,0,0), but weak at Q=(2.1,0,2) whereas the opposite is true for the LO phonon. At increased h the LA intensity drops dramatically, and the distinction between L=0 and L=2 disappears by h=0.2 (Fig. 3e). 

This behavior follows from conventional lattice dynamics. Since perovskite oxides have large unit cells containing heavy atoms, the energy of the first optic branches is relatively small. In BSCCO low-energy optic phonons disperse gradually from $\sim$4 and $\sim$7 meV (Fig. \ref{fig:Combined}b,d). As a result, the LA branch, which forms a cone at small q in the h-k plane, crosses the first optic branch around h=0.15 (Fig. \ref{fig:Combined}c and Fig. S3). Since an interaction between these branches is allowed by symmetry \cite{pintschovius_oxygen_2004, pintschovius_electronphonon_2005}, they make an avoided crossing where their eigenvectors mix and their energies are pushed apart. In addition to the crossing with an LA branch, we consider a crossing with another acoustic branch originating from the nearby superlattice reflections. In the following section we show that these branch crossings can account for the features of acoustic phonon dispersion discussed above (Fig. 2f), as well as a multi-peak profile with additional intensity appearing at low energies at h=0.2-0.3 r.l.u. (Fig. 3g).

\begin{figure}[htb!]
\includegraphics[width=0.95\linewidth]{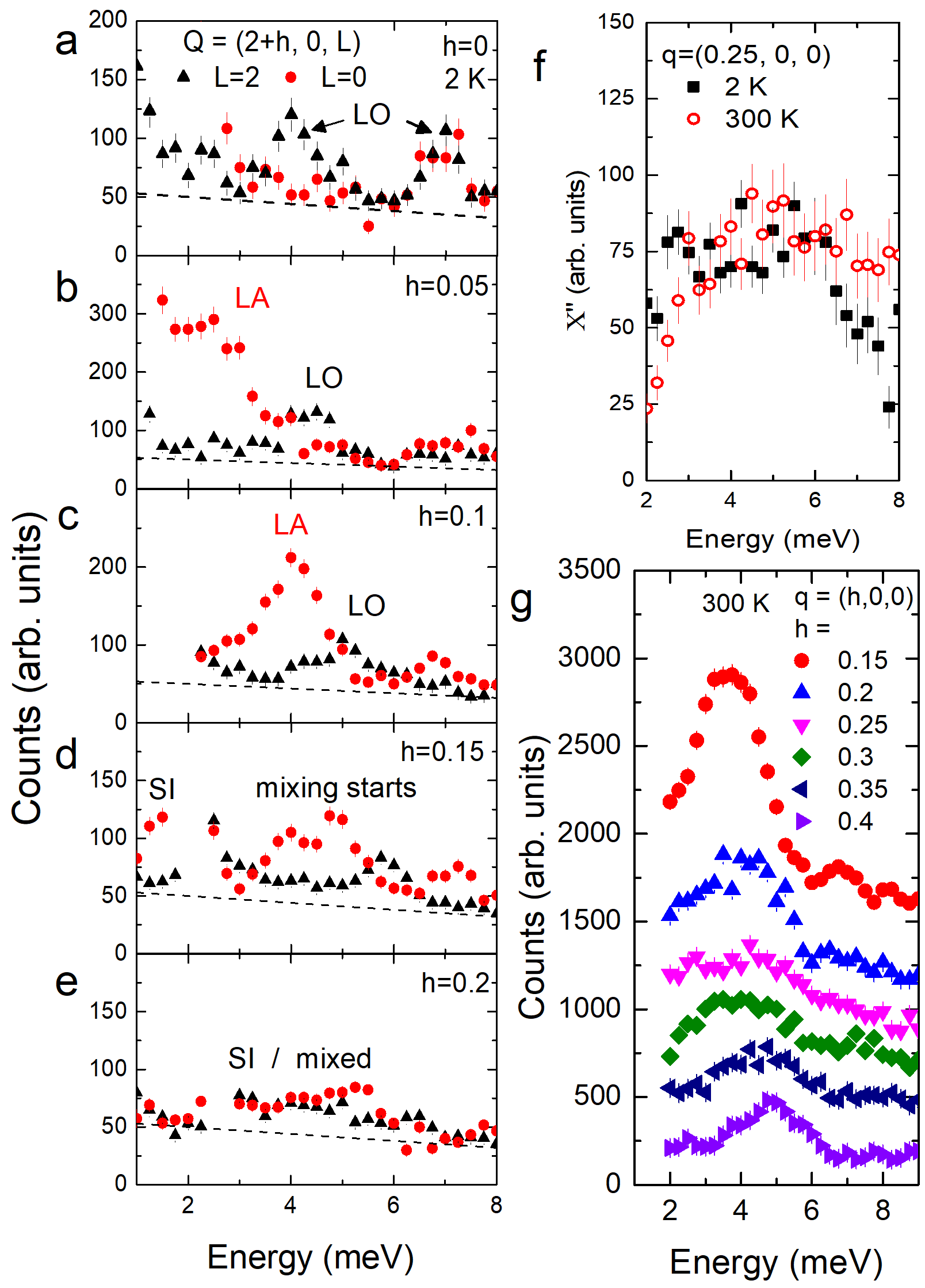}
\caption{Longitudinal phonons along the h-direction. (a-e) T=2K, E$_f$=8meV. Black triangles: total wavevector \textbf{Q}=(2+h,0,2). Red circles: total wavevector \textbf{Q}=(2+h,0,0). Dashed lines denote linear background determined from comparison of the data at different Q. SI is the supermodulation TA phonon intensity. (f) Phonon spectrum at \textbf{Q}=(2.25,0,0). Black squares: T=2K. Open red circles: T=300K. The intensity was divided by the Bose factor to obtain X$''$. E$_f$=8meV. (g) Phonon spectra at T=300K, E$_f$=14.7meV, at \textbf{Q}=(2,0,0)+\textbf{q}. Curves are offset vertically by approximately 200 points. 300K data along other directions are shown in Fig. S4}
\label{fig:lowQ_Phonons}
\end{figure}

Due to steep phonon dispersion inside the neutron scattering resolution ellipsoid, it was impossible to measure the intrinsic acoustic phonon linewidth at small crystal momentum on a standard triple-axis spectrometer. Our phonon lifetime measurements on the TRISP spectrometer at FRMII at Q=(2.11,0,0) gave a lifetime contribution to the intrinsic phonon linewidth of 217$\pm$13$\mu$eV at 20K, and 223$\pm$13.7$\mu$eV at 300K (33.3 ps) (Fig. S5). This is more than five times smaller than the intrinsic phonon linewidths measured by  another group using IXS \cite{He_Persistentlowenergyphonon-2018}. We found no change in the phonon response function across either the superconducting or pseudogap transition temperatures (e.g. Fig 3f).


\section{Discussion and Conclusions}

An interesting feature of our data is the low energy spectral weight near q$_{\text{co}}$ indicated by a shaded rectangle in Fig. 2f. It can originate from damping of the low-lying optic branch by incipient CDW fluctuations. Alternatively, it can arise from two or more closely-spaced branches that cannot be clearly resolved in the experiment as occurs during branch anticrossing.  (see for example Ref. \cite{He_Persistentlowenergyphonon-2018}). Here we argue that the latter mechanism with extra branches originating from the superstructure modulation \cite{le_page_origin_1989,hiroi_modulated_1991} (see Fig. S1) is in better agreement with experimental observations. 

The relevant superlattice reflection at Q$_S$=(2.1,0.1,$\pm$1) is of the same order of magnitude as the fundamental Q= [2,0,0] Bragg peak. Clearly observable acoustic phonons emanating from this Q$_S$ have been reported by Ref. \cite{EtrillardAcousticphononsaperiodic2001}.  

\begin{figure}[htb!]
\includegraphics[width=0.95\linewidth]{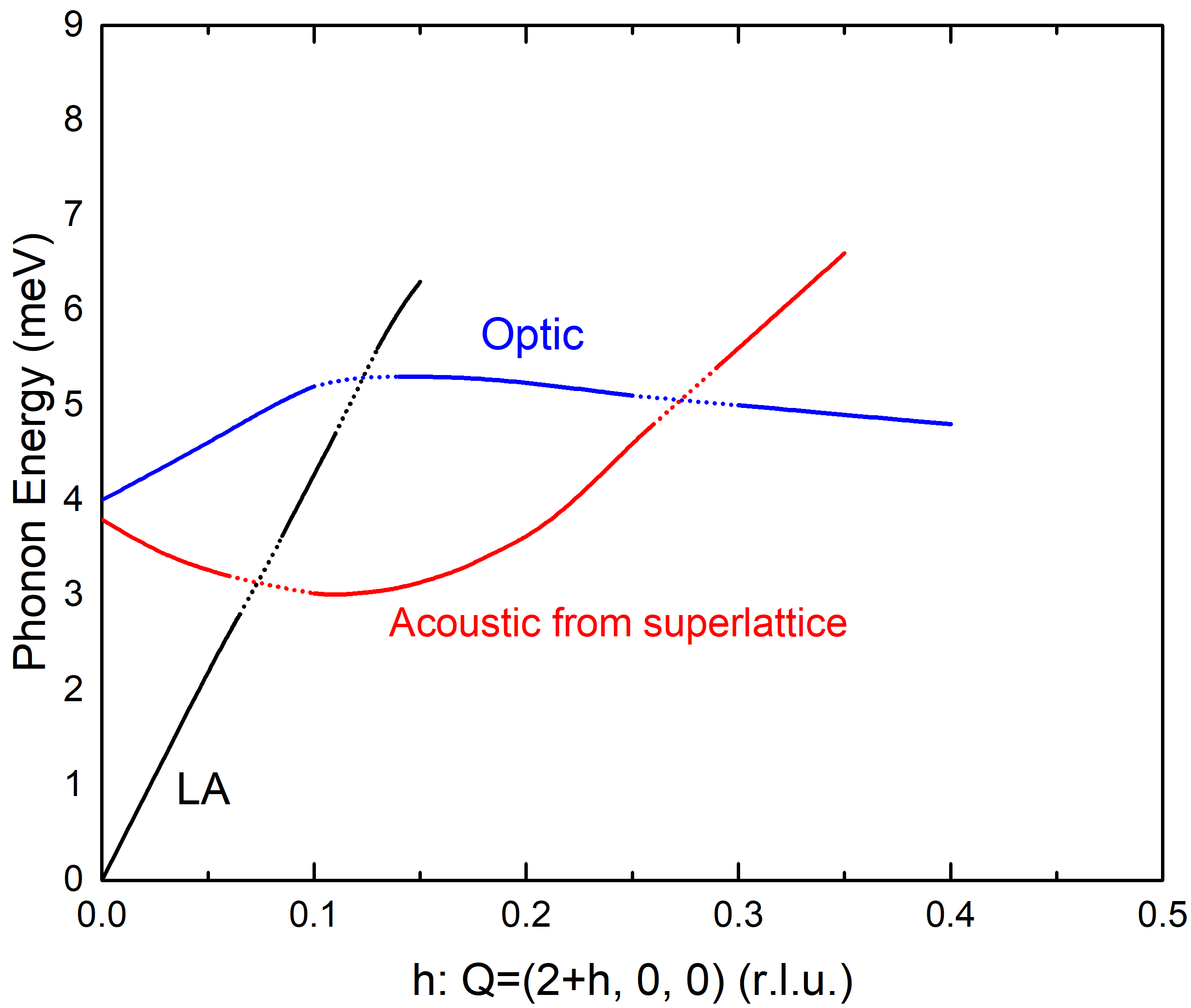}
\caption{This schematic represents how acoustic phonons originating from the superstructure modulation interfere with those originating at the (2,0,0) Bragg peak as well as with a low-lying LO branch. The black line originating at the origin shows the longitudinal acoustic (LA) phonon originating at (2,0,0) while the red line shows the acoustic phonon originating from the superstructural peak at (2.1, 0.1, $\pm$1). The dotted lines where the two phonons cross represents the possible mixing. Note that the lines represent phonon dispersions, but structure factors may be small or large at different wavevectors, and the actual situation in BSCCO is more complicated as explained in the text.}
\label{fig:Phonon_Schematic}
\end{figure}

Figure \ref{fig:Phonon_Schematic} shows the schematic of phonon dispersions along the H 0 0 direction assuming that there is an acoustic phonon branch originating at Q=(2,0,0) and another acoustic branch originating at Q$_S$=(2.1,0.1,1). Note that the second branch has a minimum energy at H=2.1 where it is closest to Q$_S$. Due to imperfect instrument resolution, the peaks from two branches will appear as one broad peak around h=0.2-0.3.

In BSCCO the situation is actually more complicated: Ref. \cite{EtrillardAcousticphononsaperiodic2001} has shown that there are two acoustic branches along the modulation Bragg peaks and one branch in the perpendicular direction. In addition, Q$_S$ does not lie on a high-symmetry direction, so it will contribute both TA and LA branches. Furthermore, there is an additional Q$_S$=(2.1,0.1,-1), which has its own acoustic phonons. This results in at least eight branches instead of two making the phonon spectrum very complicated. Disentangling it is outside the scope of the present work, but there is a high likelihood that phonon spectra in the vicinity of the crossing of all these acoustic branches with the optic branch and with each other, will be broad for reasons unrelated to incipient CDW or other electron-phonon effects.

The temperature dependence of the phonons around q$_{\text{co}}$ is not consistent with the incipient CDW instability where acoustic phonon softening is expected upon cooling \cite{bonnoit_probing_2012}. Instead the observed temperature-dependence is consistent with the Bose factor, as expected from the conventional lattice dynamical effect outlined above (see Fig. \ref{fig:lowQ_Phonons}f). Furthermore, it has been reported that the broadening disappears when the sample is doped with Pb, which destroys the superstructure \cite{bonnoit_probing_2012}. Based on all this evidence we conclude that the apparent line broadening and low energy phonon spectral weight around h=0.2-0.3 is a result of the superstructure modulation.

We now discuss the possible role of forward scattering by the small wavevector LA phonons in the low-energy electronic dispersion kink observed by ARPES. First, we show that strong electron-phonon coupling, which manifests itself in phonon lifetime broadening, does not extend far from the zone center.

The best way to isolate pure lifetime broadening from other mechanisms impacting the phonon linewidth is to measure the phonon lifetime directly as we have done for the acoustic phonon at Q=(2.11,0,0) using neutron spin-echo measurements. These gave a very long lifetime, about 5 times longer than deduced in Ref. \cite{He_Persistentlowenergyphonon-2018} from IXS measurements of linewidth. This lifetime is similar to that of the Raman phonon in undoped Si at low temperature where any electronic mechanism can be excluded.

The direct observation of a very long phonon lifetime at q=(0.11,0,0) using the spin-echo technique shows that electron-phonon coupling is already negligible by h$\sim$0.1, thus scattering of electrons by LA phonons that modulate material density in the ab-plane could be allowed only near the zone center. On the other hand, scattering of electrons by the LA branch is unconstrained for the c-axis component, because phonon dispersion along c* is shallow (Fig. \ref{fig:Combined}e), so the phase space for forward scattering by the ab-component of acoustic phonons can be large. 

Other copper oxide superconductors are characterized by low-lying optic branches that would result in similar lattice dynamics. For example this mechanism will limit large electronic phonon self-energy for the LA branch in the theory of Ref. \cite{von_oelsen_phonon_2010} to small wavevectors.

An experimental doping-dependent study of the kink in Bi-2201 and theoretical work on Bi-2212 show a significant decrease of the kink energy upon doping \cite{kondo_10meV_kink_2201_2013, johnston_acoustic_ph_2012}. However, low-energy phonons in underdoped and overdoped Bi-2212 are either doping independent or harden slightly with increased doping \cite{He_Persistentlowenergyphonon-2018}, which would make the kink doping-independent or will push it in the opposite direction. 

It has been proposed that doping-dependent Thomas-Fermi screening reduces the electron-phonon coupling strength away from small LA phonon wavevectors \cite{johnston_acoustic_ph_2012}. But the crossing between the branches limits the range of conventional electron-phonon coupling regardless of how the Thomas-Fermi wavevector evolves with doping.

These observations leave a phonon-based mechanism for the low-energy electronic dispersion kink with the problem of explaining the doping dependence. Without phonon spectra on multiple samples with various doping levels, addressing this directly is outside of the scope of the current work. However, we contend that the LA phonon forward scattering mechanism adequately explains the kink in optimally-doped BSCCO. Furthermore, small changes in the phonon crossing point with respect to q can lead to significant changes in the kink energy, as the kink energy will change with both the phonon energy $\omega_\text{ph}$(q) and the gap-referenced energy $\Delta$(q), thus the proposed mechanism may offer a way of explaining the doping dependence.

We now address the possible role of phonon forward scattering in pair forming/breaking. Fig. \ref{fig:Phonon_Schematic_1}e shows the schematic of the Fermi surface in optimally-doped BSCCO plotted together with the range in q-space where particular electrons at the center of the circles can be scattered according to our results. Aside from the region very close to the node, the LA branch scatters antinodal electrons between the parts of the Fermi surface with the same sign (within the blue circles). Based on kinematics, forward scattering at these regions will be pair-forming. Very near the node, electrons will scatter between parts of the Fermi surface where the gaps have opposite signs, which should be nominally pairbreaking, but it should have a negligible effect on pairing because it is near the node \cite{AChubukovPrivate}. Thus on balance forward scattering by the LA phonons will weakly benefit d-wave pairing. Its influence on T$_c$ is very small because the phonon energy is well below the superconducting gap. These observations suggest that low-energy phonons do not impact superconductivity in any significant way.

To conclude, we showed that the observed linewidth enhancement near the CDW ordering wavevector is not related to electronic degrees of freedom, but more likely originates from an extra low-lying acoustic phonon branch emanating from a nearby superlattice modulation reflection with wavevector Q=(2.1,0.1,$\pm$1). We also found that electron-phonon coupling to LA phonons expected from conventional theory should be limited to phonons near the zone center by branch anticrossings. This would lead to forward scattering of electrons by these phonons, which is very weakly beneficial for d-wave pairing. An acoustic phonon-based mechanism for the low-energy electronic dispersion kinks may be consistent with our results if an explanation for the previously reported doping-dependence of the kinks is found.

A.M.M. and D.R. were supported by the DOE, Office of Basic Energy Sciences, Office of Science, under Contract No. DE-SC0006939. G. G. was supported by the Office of Basic Energy Sciences (BES), Division of Materials Sciences and Engineering, U.S. Department of Energy (DOE), under Contract No. DE-SC0112704. A portion of this research used resources at the High Flux Isotope Reactor, a DOE Office of Science User Facility operated by the Oak Ridge National Laboratory. A portion of this work is based upon experiments performed at the TRISP instrument operated by MPG at the Forschungs-Neutronenquelle Heinz Maier-Leibnitz (FRM II), Garching, Germany.
\bibliographystyle{apsrev4-1}
\bibliography{bibli.bib}

\foreach \x in {1,...,6}
{%
\clearpage
\includepdf[pages={\x}]{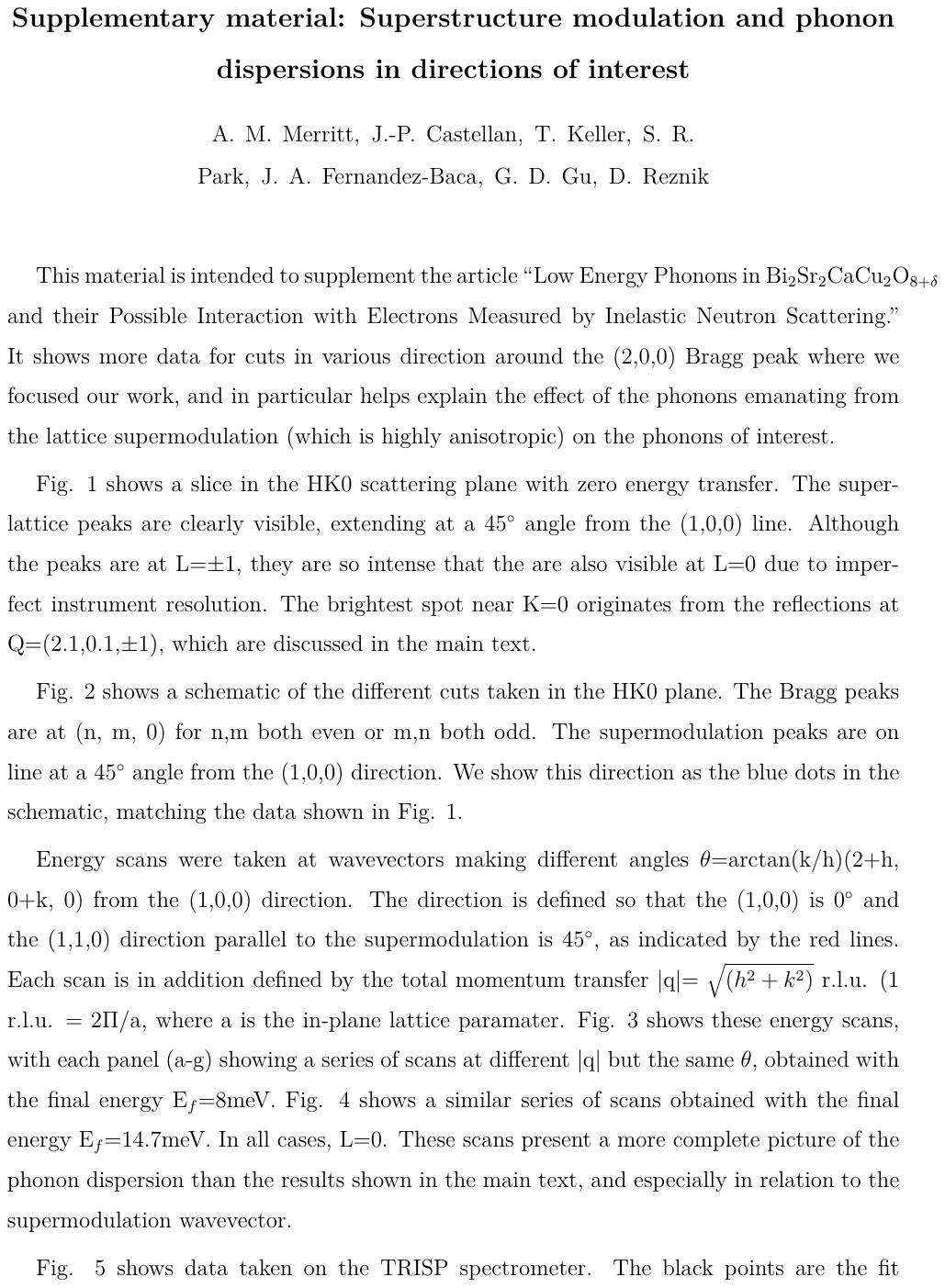} 
}

\end{document}